\documentclass[prl,twocolumn,showpacs]{revtex4}
\usepackage{epsfig}
\usepackage{amsbsy}
\usepackage{amsmath}
\usepackage{graphicx}
\usepackage{latexsym}
\usepackage{amssymb}

\begin{document}

\title{The origin of normal heat conduction in one-dimensional classics system }
\author{Haibin Li, Xiaotian Xin}

\affiliation{Department of applied physics, Zhejiang University of
Technology, Hangzhou 310023, China.}

\date{\today}
\begin{abstract}

We propose a new one-dimensional lattice model with strong
asymmetric interaction potential and investigate heat conduction
in this model numerically. We find that Fourier law is obeyed.
Based on the phonon theory, we find a new scattering mechanism of
phonon because of the breaking of the lattice segment. It is shown
that in most of scattering process in this model momentum is
destroyed as well as the Umklapp phonon-phonon scattering process
which leads to the normal heat conduction. At last, we extend our
analysis to the same class model with asymmetry interaction
potential and get a general conclusion.

\end{abstract}


\maketitle

The heat conduction in one-dimensional classic system is a long
debated problem in statistical physics. Although Peierls's
theory~\cite{Peie} shows that the Umklapp phonon scattering
process can lead to the normal heat conduction in lattice system,
but it is not definitely clear in one-dimensional system what is
the sufficient and necessary condition for the valid of Fourier
law. This problem has been attracting much attention in these two
decades~\cite{Lepri}. It is previously believed that the break of
momentum conservation may be the reason for the normal heat
conduction. The numerical studies~\cite{Casati, Zh1, Zh2} showed
that in one-dimensional system subjected to on-site potential heat
conductivity is independent on system size, which indicates the
valid of Fourier law. The momentum is not conserved in these
models because of external potential. However, such external
potential is not nature in real material, just in some treatment
of many-body system the mean field theory can produce this
potential. On the other hand, it has been suggested analytically
that the conservation of momentum is the key for the anomalous
conductivity in low dimension~\cite{Prosen}. One can consider
one-dimensional model with the mentum conservation without
external potential such as $Fermi-Pasta-Ulam(FPU)$ model. The
interaction between phonons will arise because of nonlinear term
in interaction between particles and it was expected there should
be a finite heat conductivity. However, the numerical results
showed that the heat conductivity $\kappa$ is
size-dependent~\cite{Lepri1} and divergent versus the system size
as $\kappa \sim L^\alpha$, which was also confirmed by the
mode-coupling theory~\cite{Lepri2}, kinetic theory~\cite{Perev}
and fluid hydrodynamic theory~\cite{Nara}. The studies on
one-dimensional hard-point model also showed the divergent of heat
conductivity and it is believed that in this momentum conserved
model heat conduction is anomalous and $\alpha$ is common
university of this kind of model~\cite{Hai, Grass}. However, there
are few results which are not in agreement with this belief.
One-dimensional rotor model was showed to have a finite heat
conductivity~\cite{Giar,Gend}. Recently, one-dimensional lattice
system with asymmetric interaction potential was studied
numerically and found to have a finite heat
conductivity~\cite{Zh3}.

In this paper, our goal is to find the origin of normal heat
conduction in one-dimensional system without introducing a
external potential, We investigate the heat conduction in a new
proposed one-dimensional lattice model using molecule simulation
and find Fourier law is obeyed. Using the phonon theory, we find a
new scattering process which does not occurs in one-dimensional
lattice with symmetric interaction. It will destroy the momentum
conservation of phonon and leads to the normal heat conduction.

We consider a new one-dimensional lattice system with Hamiltonian

\begin{equation}
H=\sum_n \frac{p_i^2}{2}+V(x_i-x_{i-1}-a)
\end{equation}
where $x_i$ and $p_i$ are the position and momentum of $i$th
particle respectively and $V$ is the interaction potential energy.
$a$ is the lattice constant which is the equilibrium distance
between pairs particles. We take a special form of the interaction
potential. We assume the interaction between pairs particles is
still spring like. However, when the distance between neighbor
particles is smaller than the lattice constant, there is
interaction between them. If the distance is larger than lattice
constant, the interaction vanish. Hence, this interaction
potential is
\begin{eqnarray}V(x)=
\begin{cases}
\frac{1}{2}x^2, &x<0\cr 0, &x\geq 0 \end{cases}
\end{eqnarray}
In Fig.1, we give a schematic picture of the interaction
potential. We can find this model has not only the properties of
lattice, but also of the gas model.

To investigate the heat conduction using the computer simulation,
the model should be connected with heat bath. We adapt
Noose-Hoover~\cite{Nose} heat bath on the ends of the present
model. By setting the temperature difference, we can calculate the
heat current in this model after the thermal steady state is
approached.

We find that in this model temperature gradient can be well formed
along the chain if the simulation time is long enough, which
indicates the local thermal equilibrium state can be reached. In
Fig. 2, we plot the heat conductivity $\kappa$ as the function of
the system size $L$, we note that when the system size is small,
the heat conductivity is divergent. However, when the system size
is larger than 2000, $\kappa$ become divergent slowly and comes
into a saturate value which mean we can expect a normal heat
conduction. This result shows that even in a momentum conservation
one-dimensional system, a finite heat conduction can be found.

Comparing the interaction potential used here with the interaction
potential used in ref.\cite{Zh3}, we can see although they have
different form, but they both have asymmetric properties. And, the
asymmetry in our model is much stronger than  others and heat
conductivity in our model is easy to converge. So a question
arises, is the asymmetric of interaction lead to normal heat
conduction? If so, what is the underlying mechanism. In the
following, we will investigate these problems.

\begin{figure}
\includegraphics[width=0.45\textwidth]{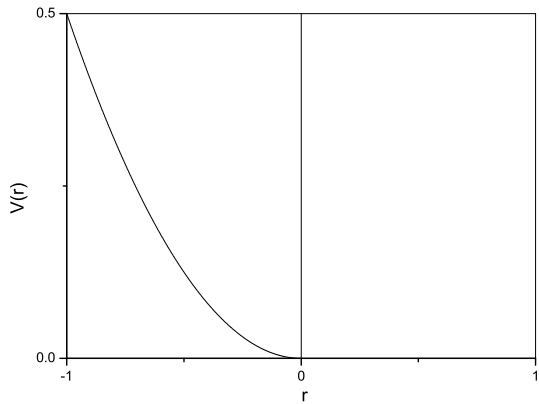}
\caption{The schematic of the interaction potential energy. }.
\end{figure}

\begin{figure}
\includegraphics[width=0.45\textwidth]{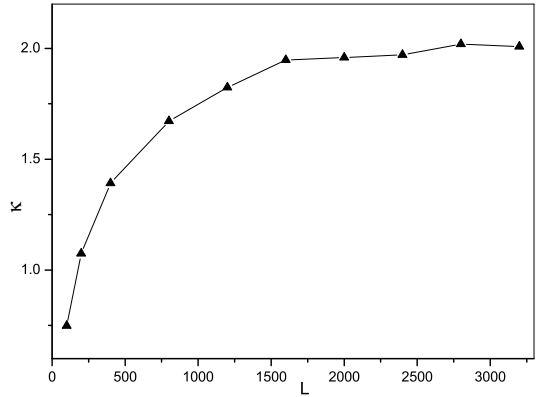}
\caption{The heat conductivity $\kappa$ as the function the system
size. The temperature at two heat bath are $T_+=0.3$ and $T_-=0.2$
respectively }.
\end{figure}
First we consider one-dimensional harmonic chain of particle each
of them has mass $m$ and is coupled each other by the interaction
$V(x)=\frac{\omega_0^2}{2}x^2$. With the period boundary condition
 a set of normal mode coordinates are given by

\begin{eqnarray}
Q_k=\frac{1}{\sqrt{N}}\sum_{l=1}^N x_l exp(ikl),\nonumber \\
P_k=\frac{1}{\sqrt{N}}\sum_{l=1}^N p_l exp(ikl),
\end{eqnarray}
where $k$ is the wave vector and given by $k=2\pi j/N$ and $j$ is
a integer chosen so that $k$ is in the Brillouin zone
$(-\pi,\pi)$. In quantum mechanic, one can use the annihilation
and creation operators $a_k$ and $a_k^+$,and $a_k$ is defined as
\begin{equation}
a_k=\sqrt {\frac{m\Omega_k}{2\hbar}}(Q_k+\frac{iP_k}{m\Omega_k}),
\end{equation}
where $\Omega_k$ are the normal mode frequencies given by
\begin{equation}
\Omega_k=2\omega_0 \mid \sin \frac{k}{2}\mid=\omega_o \omega_k,
\end{equation}
the Hamiltonian is diagonalized  as
\begin{equation}
H=\sum_{k} \omega_k (\frac{1}{2}+a_k^+ a_k).
\end{equation}
The quantum of vibrational energy is called a phonon. In the
Peierls theory of solid, the heat conduction is carried by
phonons, if there is no scattering between phonons, the mean free
path will become infinite and temperature gradient can't formed.
By introducing the nonlinear interaction between particles, it is
shown that there exists collisions between phonons. If the
momentum of phonons is conserved during the collision, this
process is called normal process and if the momentum is not
conserved it is called Umklapp process. Only the later will
contribute to the normal heat conduction because in which the heat
resistance can be formed. In the previous works on this subject,
the main aim was to find a scattering mechanism to cause normal
heat conduction. The external potential is believed to be a key to
answer this question. In real space the momentum conservation is
destroyed by the external potential, so it is suggested that it is
true for phonon.

\begin{figure}
\includegraphics[width=0.45\textwidth]{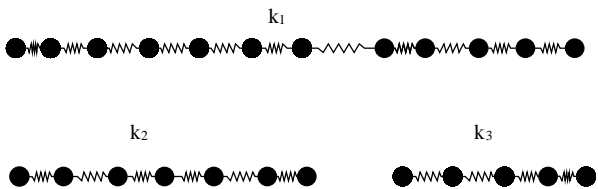}
\caption{The schematic of the breaking of lattice segment }.
\end{figure}

Let's consider our model which has the features of both the
lattice model and gas model. We can not apply phonon theory to it
directly because pair particles have no interaction if the
distance between them is larger than $a$. Assuming the system is
composed of $N$ particles and the length is $L=Na$ with the period
boundary condition, when it is in ground state that each particle
is at it's equilibrium position, there is no force between them.
When an excitation emerges along the chain, particles are pushed
from their equilibrium position. Some of them are close to its
neighbor and the interaction occurs between them. As a result, a
lattice segment is formed and between segments the distance is
larger than $a$, which means the model in thermal state is
composed by lattice segments. We can use phonon theory to describe
lattice segment.

Assuming a lattice segment composed by $N_1$ particles and its
length is $L_1=N_1 a$. The wave vectors of phonons excited in it
are
\begin{equation}
k_1=\frac{2\pi}{L_1}j_1,
\end{equation}
where $j$ is a integer and takes the values from $1$ to $N_1-1$.
In the subsequent evolution of model, it is expected that at
certain time, there will be a breaking in this segment that one
particle will move far away one of its neighbor and their distance
is larger than $a$. As a result, the original segment now splints
into two segments as shown in Fig.3. Assuming the number of
particles in these two segment are $N_2$ and $N_3$ and the lengths
of them are $L_2$ and $L_3$ respectively. Hence, the wave vectors
of phonons in these two new lattice segments are
\begin{eqnarray}\label{nk}
k_2=\frac{2\pi}{L_2}j_2, \nonumber\\
k_3=\frac{2\pi}{L_3}j_3,
\end{eqnarray}
where $j_2$ and $j_3$ are also positive integers and their maximal
values are $N_2-1$ and $N_3-1$. As $L_1=L_2+L_3$, it is easy to
see

\begin{equation} \label{ieq}
k_2 > k_1, k_3 > k_1.
\end{equation}
After the breaking of segment phonons are changed. That is to say
that the breaking of segment is a mechanism that causes the
change, namely scattering, of phonon. Let's consider the one in
original segment with the wave vector $k'_1=2\pi /L_1 $ and its
quasi-momenta is $\hbar k'_1$. In the breaking of segment, it will
split into many phonons. The wave vectors of new generated phonons
take the values in equation (\ref{nk}). Consider a simple three
phonons process that one phonon splits into two phonons, we denote
their momenta as $\hbar k''_2$ and $\hbar k'''_3$, it is easy to
find that the momentum is not conserved that

\begin{equation}
\hbar k'_1 \neq \hbar k''_2+\hbar k'''_3,
\end{equation}
due to the equation(\ref{ieq}), $k''_2$ and $k'''_3$ are larger
than $k'_1$. To make the equivalent of both sides of above
equation, one should add another momenta into left side. We denote
its wave vector by $G$ and get a equation as following
\begin{equation}
\hbar k'_1 +\hbar G =\hbar k''_2+\hbar k'''_3.
\end{equation}
This is a Umklapp process. In fact, in the breaking of segment,
the phonon $k'_1$ should split into many new phonons in two new
segments. The wave vector of each new phonon is larger than the
old one. So, it is also showed that such multi-phonon process
should satisfy equation as

\begin{equation}
\hbar k'_1 +\hbar G' =\sum_{i=1}^{m_1}\hbar
k^i_2+\sum_{j=1}^{m_2}\hbar k^j_3,
\end{equation}
where $m_1$ and $m_2$ are the numbers of phonons involved in this
process in the two new segments. $G'$ is an additional term to
fulfill the equality of equation. This process is also a kind of
Umklapp process. For other phonons in initial segment, it can also
be showed in this way that in most scattering processes the
momentum conservation are destroyed, although for the phonons with
larger wave vector in segment before breaking, there will be few
processes that satisfies momentum conservation. On the other hand,
as is well known, only the phonon with low wave vector contribute
to the heat conduction. In term of our illustration, the
scattering processes of phonons with small wave vectors are almost
Umklapp process. Hence, we can conclude that the breaking of
lattice segment in this model is a scattering mechanism of phonons
which destroys the momentum conservation. As a result we can
expect the normal heat conduction in this model confirmed by the
numerical result in Fig.2.

As our model has the asymmetric interaction which cause the
breaking in the evolution of model, next, we extend our analysis
to the models with other asymmetric interaction, such as
Lennard-Jones potential written as

\begin{equation}
V(r)=\varepsilon [(\frac{r_m}{r})^{12}-2(\frac{r_m}{r})^6],
\end{equation}
where $r$ is the distance between pairs particles. For
one-dimensional lattice, $r=x_i-x_{i-1}$. $r_m$ is the distance at
which potential has the minimum. $\varepsilon$ is the depth of
potential well. As well known, Lennard-Jones potential is
asymmetric around the minimum. To show the asymmetric more
clearly, we calculate the differential of potential around the the
minimum of potential and define a rate as

\begin{equation}
\gamma=\frac{d V(r) /d r \mid _{r=r_m+\Delta r}}{d V(r) /d r \mid
_{r=r_m-\Delta r}}
\end{equation}
where $\Delta r$ is the relative displacement of particle from its
equilibrium position. Some results are illustrated in Fig.4, we
can find that for fixed parameters of potential the differential
of potential at the same displacement on the two sides of minimum
become more different when the displacement $\Delta r$ becomes
large, which shows the asymmetric of interaction potential. And,
we find that $r_m$ determines the degree of asymmetry of
potential. When $r_m$ becomes small, the differential of
potential, namely, the force acted on pairs particles becomes much
smaller when the two particles leave apart than the case when two
particles move close to each other.

\begin{figure}
\begin{center}
\includegraphics[width=0.45\textwidth]{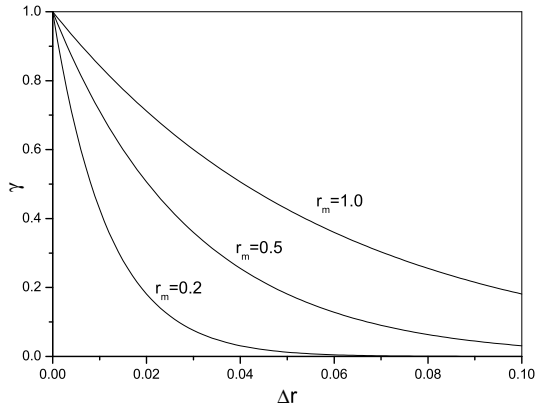}

\caption{The rate $\gamma$ versus the relatively difference of
displacement from the minimum $\Delta r$ in the Lennard-Jones
potential, $\varepsilon =1.0$. }.
\end{center}
\end{figure}

To use phonon theory, the general treatment of nonlinear potential
is to approximate it by expanding it around the minimum and just
take the second order term as the first order term is zero.
However, as the expansion coefficient of second term, $d V /d r$
is taken at the minimum no matter what sigh of $\delta r$. Hence,
the asymmetry is lost in the approximation.

To involve the asymmetry in the linear approximation, we expand
the potential along the trajectory. We can get the function of
linear term. For simplicity, we use the average linear term
function. And the form of it on the left side and right side of
minimum denoted by $k_l$ and $k_r$ are defined as following
\begin{eqnarray}
k_l=\frac{1}{r_m-r_l}\int_{r_l}^{r_m}\frac{d V(r)}{d r}dr=\frac{V(r_m)-V(r_l)}{r_m-r_l} \nonumber \\
k_r=\frac{1}{r_r-r_m}\int_{r_m}^{r_r}\frac{d
V(r)}{dr}dr=\frac{V(r_r)-V(r_m)}{r_r-r_m}
\end{eqnarray}
Considering the case $r_m-r_l=r_r-r_m$, it is easy to conclude
from Fig.4 that $k_l > k_r$, showing the asymmetry. Thus, we can
consider the evolution of this model in linear approximation. The
coupling constant between pairs particles are dependent on
distance, taking value $k_l$ or $k_r$. In other words, it is
time-dependent. In the evolution, with the change of distance of
pairs particles, there will be a crossover at which the couple
constant will switch between $k_l$ and $k_r$, then the phonon
spectrum will change. It is a scattering mechanic of phonons.
Consider the small $r_m$, it can be expected that $k_l\gg k_r$, we
can ignore $k_r$ and only $k_l$ exists, then the model with
Lennard-Jones potential is reduced to our model. Hence our finding
of the scatting of phonon can be applied to such model and the
same conclusion is obtained.

In conclusion, we found normal heat conduction in our
one-dimensional lattice model with asymmetric interaction, that
provides another example which obey Fourier law with momentum
conserved. It is to say that the momentum conservation is not the
necessary condition for normal heat conduction. Furthermore, we
found in our model, due to the asymmetric interaction potential,
there is a scattering mechanism caused by the breaking of the
segment, which leads to the normal heat conduction. That is to say
the phonon spectrum is time-dependent. We can apply our analysis
to other model with asymmetric interaction. As is well known, the
asymmetric interaction potential is the reason for the expansion
of solid potential~\cite{Patter}. Our finding indicates that it
can account for the normal heat conduction.

\end{document}